\documentstyle{article}
\begin{document}
\LARGE
\begin{center}
\bf Causality, Time Arrow and Half Cycling Universe

\vspace*{0.4in}
\normalsize \large \rm 

Wu Zhong Chao

Dept. of Physics

Beijing Normal University

Beijing 100875, China

(April 10, 1998)

\vspace*{0.6in}
\large
\bf
Abstract
\end{center}
\vspace*{.1in}
\rm
\normalsize

If one introduces causality into quantum cosmology, then the
prescription for the no-boundary universe should be
revised. We show that the thermodynamic arrow of time associated
with the perturbation modes should be reversed at the maximum
expansion for the oscillating Hawking model. To an observer
equipped with the time arrow, the universe will terminate its
evolution after an half cycle. 

\vspace*{0.8in}

PACS number(s): 98.80.Hw, 98.80.Bp, 04.60.Kz, 04.70.Dy


\vspace*{0.8in}

e-mail: wu@axp3g9.icra.it $\;\;\;\;$

\pagebreak

In the no-boundary universe [1], the Hawking model has been
extensively investigated [2]. It is a closed $FRW$ universe
coupled to a scalar field $\phi= \phi (t)$ with potential
$V(\phi) = m^2
\phi^2$. Its
Euclidean metric is described by
\begin{equation}
ds^2 =  d\tau^2 + b^2(\tau)(d\chi^2 + \sin^2\chi (d\theta^2 +
\sin^2\theta d\psi^2)).
\end{equation}

The Euclidean action is
\begin{equation}
\bar{I} = - \frac{3\pi}{4} \int d\tau b\left ( \dot{b}^2 + 1 -
\frac{b^2\dot{\phi}^2}{3} - \frac{m^2 b^2 \phi^2}{3} \right ),
\end{equation}
where dots denote derivatives with respect to the imaginary time
$\tau = it$ and the scalar field has been rescaled by multiplying
the factor of $(4 \pi)^{\frac{1}{2}}$ for convenience.

The field $\phi$ and the scale $b$ obey the equations
\begin{equation}
\ddot{\phi} + \frac{3\dot{b}\dot{\phi}}{b} = m^2 \phi,\;\;\;
\ddot{b} = - \frac{2b\dot{\phi}^2}{3} - \frac{m^2b \phi^2}{3}
\end{equation}
and the Hamiltonian constraint
\begin{equation}
\frac{\dot{b}^2}{b^2} - \frac{1}{b^2} - \frac{ \dot{\phi}^2}{3} +
\frac {m^2 \phi^2}{3} = 0.
\end{equation}

The quantum state of the universe is defined by the path integral
over all compact 4-manifolds with the configuration of the wave
function, the 3-geometry and matter field on it, as the only
boundary. The wave function can be approximated by the
exponential of the negative of the action of the complex solution
with the boundary. This is called the  $WKB$ level. We shall work
at this level in this paper. If one uses a Euclidean solution,
then
it can be  described by a deformed 4-sphere. Its south hemisphere
can be approximated by a half 4-sphere and the scalar field
increases
slowly from its south pole. As soon as the 4-sphere reaches its
maximum size at the equator, 
the scalar field will increase rapidly and then the manifold will
collapse into a singularity. 

For a general complex solution, the regularity conditions at the
south pole $\tau_S $ implied by the no-boundary proposal are 
\begin{equation}
b =0, \;\;\; \dot{b} = 1,\;\;\;  \phi = \phi_0, \;\;\;
\dot{\phi} = 0.
\end{equation}
The singular manifold is parametrized by the initial value
$\phi_0 $ there.

A regular Euclidean solution is called an instanton. It has been
shown that there does not exist any nontrivial regular instanton
in this model [3][4]. Equivalently, one cannot find a
compact
Euclidean regular solution with a 3-geometry (the equator) as the
only boundary on which the second fundamental form vanishes and
the normal derivative of the matter field is zero. At best, one
can only find a Euclidean solution with approximately vanishing
momenta at the equator [5]. Even when one relaxes the solution to
be complex, the situation associated with the singularity will
not change.

One way out of the trouble caused by this singularity behavior in
the
scalar model is to reinterpret the Euclidean  solution to the
field equation as a constrained gravitational instanton [4][6].
The south hemisphere of the manifold is the stationary action
solution under the condition that the 3-geometry is given, at the
maximum size where the quantum transition is supposed to occur. 
The whole manifold is made by joining this south hemisphere and
its oriented reversal as the north hemisphere. One can
also use $\phi_0$ to parametrize the 3-geometry. The variational
calculation shows that the stationary action solution should be
regular and satisfy the field equations everywhere,  with the
only possible exception at the 3-geometry equator. Therefore the
joint manifold has a stationary action under the restriction
imposed at the equator or for a fixed $\phi_0$ and qualifies as a
constrained gravitational instanton. The constrained instanton
can be used as the
seed for the creation of the universe. For this model, the
canonical momentum of the scalar field at the equator, i.e, its
normal derivative, is allowed to be nonzero. However, the nonzero
real part of the momentum conjugate to the scalar in imaginary
time is identified as  its nonzero imaginary part in real time,
therefore
it can not lead to a real evolution in the Lorentzian regime.

The Lorentzian evolution of the model can be obtained through an
analytical continuation at the equator. As mentioned above, in
order to obtain a real evolution in the Lorentzian regime with
real time one has to impose the condition at the equator that the
imaginary parts of all fields and the real parts of all conjugate
momenta in imaginary time are zero [7]. For the Hawking model
this is
\[
Im(b) = 0,\;\;\; Im(\phi) = 0,\;\;\;
\]
\begin{equation}
Re(\dot{b}) = 0,\;\;\; Re(\dot{\phi}) = 0.
\end{equation}

Apparently, to meet condition (6), one has to
find a complex solution or complex constrained instanton to
replace the Euclidean south hemisphere. One has two degrees
of freedom at  the south pole, that is $\phi_0^{Re} +
i\phi_0^{Im}$, for the complex solution, and in addition to these
one
has two more degrees of freedom for the location of the equator
in the complex time plane. At the mean time, there are 4
conditions in Eq.(6) at the equator, therefore there are
at most a discrete set of initial values of $\phi_0$ leading
to Lorentzian evolutions. We set $\tau_1 =0$ at 
the equator and then at the south pole $\tau_S = - \tau_0^{Re} +
i \tau_0^{Im}$.
Since the Lorentzian condition (6) at the equator is sufficient
to guarantee the evolution along the line $\tau^{Re} = 0$ to be 
real, then at any point on this line, condition
(6) is satisfied. One can further choose the point where
$Im(\dot{b}) = 0$ as the equator. It means that at the creation
the initial expansion rate is zero. This conclusion will
not be changed even if one includes more matter fields into the
model. 

To find the complex solutions leading to purely Lorentzian
evolutions is interesting. But this is not the focus of this
paper. Since there does not exist any  regular instanton, at the
equator the imaginary parts $Im(\dot{\phi})$ will not vanish.
This means that in the Lorentzian regime the  initial time
derivative of
the scalar at $\tau_1$ is nonzero. However, one can select a
suitable value $\phi_0$ such that $Im(\dot{\phi})$ is very small.
After the creation the universe will undergo an
inflationary period. Then the scalar $\phi$ decreases
significantly and starts to oscillate.
The inflationary period is succeeded by the matter-dominated
phase of the big bang model. The universe will reach the maximum
size at $\tau_2 = 0- i\tau_2^{Im}$ and then recollapse.

In this paper we are particularly interested in the oscillating
universe. When the universe reaches the maximum size at $\tau_2$,
it will meet condition (6) again. If at the maximum expansion the
scalar field satisfies $\phi =0 $ or $\phi^\prime =0$, where
prime
denotes the derivative with respect to real time, then the
contraction phase of the universe is exactly the time reversal of
the expansion phase. For the first case the only difference is
that the sign of the scalar should be changed. After the universe
shrinks into the initial  size, it will bounce back, and
oscillate $ad \;\;\; infinitum$. The oscillating condition can be
realized by adjusting the mass parameter $m$, in addition to the
four parameters mentioned above.

There are scalar and tensor perturbation modes around the
minisuperspace background. The tensor modes describes  the
primordial gravitational waves. They are
\begin{equation}
\delta g^T_{\mu \nu} =\sum_n b^2 \bordermatrix{ &  & \cr
                                  & 0 & 0  \cr
                                  & 0 & 2d_n G^n_{ij} \cr} ,
\end{equation}
where $G^n_{ij}$ are the transverse traceless tensor harmonics 
and $d_n$ are the amplitudes.

From the Wheeler-DeWitt equation one can derive [8]
\begin{equation}
i\frac{\partial \Psi^{(n)}_T}{\partial t} = \frac{1}{b^3} \left
[- \frac{\partial^2}{\partial d^2_n} + d^2_n (n^2 - 1) b^4 \right
] \Psi^{(n)}_T,
\end{equation}
where $\Psi^{(n)}_T$ is the wave function for the mode. This is
the Schroedinger equation for an oscillator with a
time-dependent frequency $\nu_n \approx n/b$. 

The classical evolution obeys
\begin{equation}
d^{\prime \prime}_n + \frac{2b^\prime d_n^\prime}{b} + (n^2 -
1)d_n = 0
\end{equation}
or
\begin{equation}
(bd_n)^{\prime \prime} + \left ( n^2 - 1 - \frac{b^{\prime
\prime}}{b} \right ) bd_n = 0,
\end{equation}
where, and for the rest of the paper, the prime denotes the time
derivative with respect to Lorentzian conformal time $\eta$
defined by $d\eta=b^{-1} dt$.

Now we try to work out the quantum state of the tensor mode using
the no-boundary proposal. At the moment $\tau = 0$, one has the
Lorentzian condition $\dot{a_n} = 0$ and $Im(a_n)=0$. These
conditions can be satisfied by the choice of complex values of
$a_n$ and $\dot{a_n}$  at the south pole. The regularity
condition at the south pole is $b^2d_n = 0$. This condition can
be realized by evolving Eq. (10) towards the south pole. Since
one has $b = (\tau - \tau_S)+ \lambda (\tau - \tau_S)^3 + ...
(\lambda = const. )$ at the south pole, then the potential
barrier is finite there, so $bd_n$ must be finite and $b^2d_n$ is
equal to zero at $\tau_S$. In the Euclidean regime the term
$b^{\prime \prime}/b$ can be ignored if $n
\gg \dot{b}$. This implies that the complex solution is $d_n
\approx d_n(0) \cosh(b^{-1}n\tau)$. It is noted that the
regularity condition of
the modes at the south pole in Ref. [8] is too strict. At $\tau
=0$, taking account of the fact that
the initial expansion rate is zero, one can obtain the wave
function of the mode at the $WKB$ level as in [8]
\begin{equation}
\Psi^{(n)}_T \approx \exp(-\frac{1}{2} nb^2d^2_n).
\end{equation} 
That is, the no-boundary condition implies that  each mode
initially takes the minimum excitation state allowed by the
Heisenberg uncertainty principle in quantum mechanics. 

However, when the wave length of the mode becomes equal to the
horizon scale, then the adiabatic approximation breaks down. The
wave function will then freeze. Quantum mechanically, the state
is described by a squeezed vacuum, while classically the
oscillator undergoes a parameter amplification.  After the
mode reenters the horizon in the matter dominated era, the
adiabatic approximation  ($n \gg \dot{b}$) will be recovered.
However, the
oscillator then is in a highly excited mode. It was argued that
at the
expansion phase the entropy associated with the perturbation
modes increases, and the thermodynamic arrow of time is implied
by the cosmological arrow which is defined by the direction of
the expansion of the universe [9].

Since the operator in the right hand side of Eq. (8) is time
symmetric, the amplitude of the wave function with initial value
(11) is an even function of
time at the creation moment. For the oscillating model, if we
identify the two ends of the Lorentzian time in a cycle, the
amplitude of the wave function will remain symmetric about the
identified time origin. The amplitude evolved forward from the
big bang and that evolved backward from the big crunch will be
equal at the maximum
expansion. It is noted the Schroedinger equation is of first
order in time derivative. If one follows the evolution from the
big bang until
the big crunch, he will find that the wave function will return
to the initial state at the crunch end up to a phase.

The classical equation (9) can be identified as that for a
particle
tunneling through a potential barrier $1 - n^2 + b^{\prime
\prime }/b$ in one dimensional quantum mechanics with coordinate
$\eta$. In general, the solutions are not symmetric about the
maximum expansion due to reflection of the potential barrier.
However, the no-boundary wave function $\Psi^{(n)}_T$ should be
the quantum counterpart of a classical orbit which is symmetric
about the maximum expansion. The time arrow is determined by the
direction of increasing entropy and the highly excited
perturbation modes are of higher entropy, therefore the arrow of
time should be reversed at the maximum expansion.

The scalar perturbation of the matter is
\begin{equation}
\delta \phi = \sum_n \frac{1}{\sqrt{6}}f_n Q^n,
\end{equation}
where $Q^n$ are scalar harmonics. The scalar perturbation of the
metric is
\begin{equation}
\delta g^S_{\mu \nu} =\sum_n \frac{b^2}{\sqrt{6}} 
\bordermatrix{ &  & \cr
                                  & -2g_nQ^n & k_nP^n_i  \cr
                                  & k_nP^n_i & 2a_n \Omega_{ij}
Q^n + 6b_nP^n_{ij} \cr} ,
\end{equation}
where $P^n_i = Q^n_{|n}/(n^2 - 1)$ and $P^n_{ij} =
\Omega^n_{ij}Q^n/3 + Q^n_{|ij}/(n^2 -1)$.
$k_n$ and $g_n$ are Lagrange multipliers and
induce two constraints. Therefore, there is only one true scalar
degree of freedom. In the $b_n = k_n = 0, g_n = - a_n$ gauge, one
has the constraint [10]
\begin{equation}
a^\prime_n + \frac{b^\prime}{b} a_n = -3 \phi^\prime f_n
\end{equation}
and the decoupled equation of motion for $a_n$ [10]
\begin{equation}
a^{\prime \prime}_n + 2 \left [ \frac{b^\prime}{b} -
\frac{\phi^{\prime \prime}}{\phi^\prime} \right ] a^\prime_n
+ \left [2 \left (\frac{b^\prime}{b} \right )^\prime -
\frac{2b^\prime \phi^{\prime \prime}}{b \phi^\prime} + (n^2 + 3)
\right ] a_n = 0,
\end{equation}
which can be rewritten as the equation for $\tilde{a}_n \equiv
ba_n/\phi^\prime$
\begin{equation}
\tilde{a}_n^{\prime \prime} + 
\left
[ -\frac{b^{\prime \prime}}{b} + 2 \left ( \frac{b^\prime}{b}
\right )^\prime + \left (\frac{\phi^{\prime \prime}}{\phi^{\prime
2}} \right )^\prime \phi^\prime + (n^2 + 3) \right
]\tilde{a}_n=0.
\end{equation}

Again, this takes the form for a particle tunneling through a
potential barrier in one-dimensional quantum mechanics. The
quantity $\tilde{a}_n$ behaves like the wave function.
Form (16) is very useful, since the operator within the
square bracket is of even order in the time derivative, and the
dissipation effect disappears. It can be continued into the
Euclidean regime by the Wick rotation. The Lorentzian condition
for $\tilde{a}_n$ at the equator is $\dot{\tilde{a}}_n = 0$ and
$Im{\tilde{a}_n} = 0$. For the case $n \gg \dot{b}$, all terms
in the square bracket of Eq. (16) are negligible in comparison
with $(n^2 + 3)$ in the Euclidean regime. One can use a similar
method to obtain the complex solution to meet the Lorentzian
condition at the equator and the regularity condition at the
south pole as in the case of  the tensor modes.

The above solutions can be used for construction of the wave
function for the scalar mode at the creation moment and obtaining
the minimum excitation state similar to Eq. (11).  For
simplicity, one uses the oscillating
model with $\phi = 0$ at the maximum expansion. By
identifying the big bang and the big crunch ends  Eq. (16)
becomes
symmetric about the creation moment. The
Schroedinger equation corresponding to its classical counterpart
(16) can be written. Then we use the same argument as for the
tensor modes to get a wave
function with time symmetric amplitude at the minimum size and
the maximum
expansion. Therefore, the time arrow associated with the scalar
mode also flips at the maximum expansion.
 
From the no-boundary philosophy, the wave function for the
perturbation modes is a function of the configuration only and
should be independent of the orbit of the
minisuperspace background. In this way, the quantum state of the
modes at the big bang end should be identical to that at the big
crunch end. However, when one uses analytic continuation at the
$WKB$ level, one may worry that the quantum state of the modes
may depend on the contour.  However, it turns out that this is
not the case. 

In summary, for both the tensor and scalar cases, the
perturbation modes start as a minimum excitation at the big
bang end, it will return to the same state at the crunch end. The
entropy will increase when the modes evolve from the minimum
excitation to the highly excited state. Thus, the thermodynamic
and psychological arrows of time should be flipped at
the maximum expansion [9].

It is noted that the appearance of singular terms encountered in
Eq. (16) is superficial. It has no effect in Euclidean regime. In
the Lorentzian regime, all these singular behaviors must be
cancelled or one can simply integrate the equation in the complex
time plane along a very close time line parallel to the
Lorentzian contour and then take a
limit to approach the true Lorentzian evolution.

 Our conclusion about the scalar modes is different from Ref.
[10]. The main reason is that the classical Lorentzian evolution
can be described by the wave function of a particle penetrating 
through the potential barrier. If one lets
the wave function at the big bang end be a pure propagating wave
as in [10], then at the other end, it must be a
superposition of propagating and reflecting waves. There is no
surprise that the amplitude at the crunch end becomes much
larger! 

To set the pure propagating wave at the big bang end is wrong.
First, the complex values of $d_n$ and $a_n$ are unphysical.
Second, and more importantly, for the whole perturbation
calculation leading to (8) one has not taken account of 
causality condition. The naive analytic continuation leads to
complex $d_n$ and $a_n$. In quantum field theory one has to
decompose the field into the positive and negative frequency
parts associated with creation and annihilation operators [11].
The Lorentzian  condition complies with the causality condition.
This requirement has no effect on the
background calculation, since one ignores  causality in the
minisuperspace calculation anyway.

To recover causality, one has to take the real part of the whole 
wave
function $Re(C\exp iS)$, then one obtains a couple of complex
conjugate Schroedinger equations for perturbation modes, they are
for
negative and positive frequency parts, respectively. The
Hermitian property of the perturbation field leads to real values
of $d_n$ and $a_n$. This phenomenon was previously
encountered as dealing with the perturbation modes around the
Schwarzschild-de Sitter background [12]. The modified 
prescription will leave the prediction on the origin of the
structure in the universe intact, but it does change the issue
about the time arrow in the universe. Indeed, if one chooses the
pure propagating wave at the big bang end, one has implicitly
introduced the time arrow by hand from the beginning.

The universe may keep oscillating after the big crunch. However,
to an observer equipped with the thermodynamic or psychological
arrow of time, he will find that the evolution will terminate at
the maximum expansion.  The contraction phase of the universe is
identical to its expansion phase. One ends with a half-cycling
universe [13]. 

For non-cycling Lorentzian evolutions, the above argument is no
longer valid. As we know, one has to fine tune the parameters to
yield the oscillating models. However, it is believed that  the
evolutions of the
background and the quantum state of the modes are not so
sensitive to these
parameters before the anti-inflationary period near the big
crunch.
Therefore, for the general Lorentzian orbits the time
arrow should be reversed at the maximum expansion as well, it
might even
be flipped again near the big crunch if the universe recollapses
into a
true singularity. However, even for an eternal observer, he would
not be able to cross the maximum expansion to experience this.
It is noted
that, since the entropy reaches its maximum at the maximum
expansion, the time arrow becomes very vague then.

One can even present a conjecture: under the regularity condition
at the south pole and the Lorentzian condition at the equator,
all Lorentzian orbits of the background are oscillating or
quasi-oscillating. Then what we argued for the oscillating orbits
about the time arrow applies to all orbits. But I cannot
offer a proof.

\bf References:

\vspace*{0.1in}
\rm

1. J.B. Hartle and S.W. Hawking, \it Phys. Rev. \rm \bf D\rm
\underline{28}, 2960 (1983).

2. S.W. Hawking, \it Nucl. Phys. \rm \bf B\rm \underline{239},
257 (1984).

3. A. Vilenkin, hep-th/9803084.

4. Z.C. Wu, hep-th/9803121.

5. R. Bousso and S.W. Hawking, \it Phys. Rev. \rm \bf D\rm
\underline{52}, 5659 (1995).

6. Z.C. Wu,  \it Int. J. Mod. Phys. \rm \bf D\rm\underline{6},
199 (1997); \it Gene. Rela. Grav. \rm\underline{30}, 115
(1998). 

7. N. Turok and S.W. Hawking, hep-th/9803156. $\;$ R. Bousso
and S.W. Hawking,(in preparation).

8. J.J. Halliwell and S.W. Hawking, \it Phys. Rev. \rm \bf D\rm
\underline{31}, 1777 (1985).
 
9. S.W. Hawking, \it Phys. Rev. \rm \bf D\rm \underline{32}, 2489
(1985).

10. S.W. Hawking, R. Laflamme and G.W. Lyons, \it Phys. Rev. \rm
\bf D\rm \underline{47}, 5342 (1993). 

11. S. Weinberg, \it The Quantum Field Theory, \rm (Cambridge
University Press, 1995).

12. Z.C. Wu, gr-qc/9712066.

13. Z.C. Wu, unpublished (1993).

\end{document}